\documentclass[9pt]{article}
\usepackage[letterpaper,left=6.9cm,right=1.9cm,top=3.1cm,bottom=2.3cm]{geometry} 

\input{Styles_ACCEFN.sty}  
\vspace{0.3cm} 
{\title{\fontsize{14}{14}\textbf{Radiative neutrino masses in an abelian extension of the Standard Model} \\[0.2cm]
\textcolor{gray}{\textbf{Masas radiativas de neutrinos en una extensión abeliana del Modelo Estándar}\\[0.2cm]}}
}

\author[1,*]{\fontsize{9}{9} \orcidlink{0000-0001-5617-0138} \textbf{S.F. Mantilla}}
\author[2]{ \orcidlink{0000-0002-4678-2197} \textbf{R. Martinez}}
\author[2]{\orcidlink{0009-0008-3579-0256} \textbf{J. Muñoz}}

\affil[1]{\fontsize{8}{8}Universidad Nacional Abierta y a Distancia (UNAD), Escuela de Ciencias Básicas, Tecnología e Ingeniería (ECBTI), Sede Nacional José Celestino Mutis, Calle 14 Sur 14-23, Bogotá, Colombia.}
\affil[2]{Departamento de Física, Universidad Nacional de Colombia, Ciudad Universitaria, K. 45 No. 26-85, Bogotá D.C., Colombia.}
\begin{document}

\maketitle
\section*{Abstract}
\justifying
The seesaw mechanism is the preferred methodology to obtain light neutrino masses by the introduction of a Majorana mass matrix. Moreover, radiative corrections can be done in order to improve the predictions of the model. However, when such a Majorana mass matrix has no inverse, radiative corrections turn out as the only mechanism to get massive neutrinos.  In this letter, it is shown a nonuniversal abelian extension $\mathrm{U(1)}_{X}$ of the Standard Model in which neutrinos acquire masses completely by radiative corrections through one-loop self-energies involving heavier right-handed neutrinos. Moreover, one right-handed neutrino $\nu_{R}^{2}$ also gets mass through radiative corrections involving Majorana fermions. Finally, some limit cases are drafted depending on the hierarchy among vacuum expectation values and Majorana masses.
.\\[0.2cm]
\textbf{Keywords:} Flavor Problem, Neutrino Physics, Extended Scalar Sectors, Beyond Standard Model, Fermion masses, Inverse SeeSaw Mechanism.

\section*{Introduction}
Since the very beginning of the search on particle physics, neutrinos have shown really special behaviors in comparison with other fundamental particles such as electrons, muons or quarks. Neutrino oscillations points one of their most important features, which is that they have really small but finite masses with only a left-handed chirality \cite{esteban2017updated,nufit}, despite the Standard Model (SM)\cite{glashow1961partial,salam1968elementary,weinberg1967model} demands the existence of both left and right to generate masses through the electroweak spontaneous symmetry breaking (SSB). 


In this way, the most preferred scheme to obtain small neutrino masses is the well-known \textit{seesaw mechanism} (SSM)\cite{minkowski1977mu,gell1979r,yanagida1980horizontal,mohapatra1980neutrino,schechter1980neutrino,das2017heavy,sierra2011importance,grimus2002one,das2017enhanced} which proposes the existence of right-handed neutrinos as Majorana fermions. Since they are SM singlets, right-handed neutrinos are able to present Majorana masses whose scale is fixed in order to get the suited mass scales of neutrino masses according to neutrino oscillation data. However, the simple SSM requires Majorana masses for the right-handed neutrinos at $10^{14}$ GeV, near to the usual grand unification scale and evidently out of any experimental probe available today. In order to solve this unpleasant problem, the \textit{inverse seesaw mechanism} (ISS) introduces a two-folded spectrum of right-handed neutrinos and consequently the Majorana mass scale can be lower by the addition of an additional scale at which right-handed neutrinos acquires mass\cite{mohapatra1986mechanism,mohapatra1986neutrino,hernandez2013radiative,dev2012minimal,hernandez2016predictive,catano2012neutrino,hernandez2014lepton,dias2012simple,hernandez2015s_3,das2017generation}. 

One realization of the ISS is presented in Ref. \cite{mantilla2017nonuniversal}. The particle content includes the three lepton doublets of the SM $\ell_{L}^{\alpha}$ ($\alpha=e,\mu,\tau$), three right-handed Dirac neutrinos $\nu_{R}^{\alpha}$ and three right-handed Majorana fermions $\mathcal{N}_{R}^{\alpha}$ (see Tab. \ref{tab:Leptonic-sector}). After the SSB the neutral sector gets the mass matrix $\mathbb{M}_{N}$ in the flavor basis $\mathbf{N}_{L}=(\nu^{\alpha}_{L},{\nu^{\alpha}_{R}}^{C},{\mathcal{N}^{\alpha}_{R}}^{C})$
\begin{equation}
\mathbb{M}_{N} = 
\left(\begin{array}{c c c}
0	&	\boldsymbol{\mathcal{M}}_{\nu}^{\mathrm{T}}	\\
\boldsymbol{\mathcal{M}}_{\nu}	&	\boldsymbol{\mathcal{M}}_{\mathcal{N}}
\end{array}\right) = 
\left(\begin{array}{c c c}
0	&	\mathcal{M}_{\nu}^{\mathrm{T}}	&	0	\\
\mathcal{M}_{\nu}	&	0	&	\mathcal{M}_{\mathcal{N}}^{\mathrm{T}}	\\
0	&	\mathcal{M}_{\mathcal{N}}	&	{M}_{\mathcal{N}}
\end{array}\right),
\end{equation}

Performing the inverse SSM together with the assumptions $\mathcal{M}_{\mathcal{N}} \gg \mathcal{M}_{\nu}\gg {M}_{\mathcal{N}}$, we find that $\mathbb{M}_{N}$ can be diagonalized by blocks in the following way
\begin{equation}
\left(\mathbb{V}_{L,\mathrm{SS}}^{N}\right)^{\dagger}\mathbb{M}_{N} \mathbb{V}_{L,\mathrm{SS}}^{N} = 
\begin{pmatrix}
m_{\nu}	&	0	&	0	\\
0	&	m_{N}	&	0	\\
0	&	0	&	m_{\tilde{N}}
\end{pmatrix}
\end{equation}
spanned in the mass basis $\mathbf{n}_{L}=(\nu^{i}_{L},N^{i}_{L},\tilde{N}^{i}_{L})$ ($i=1,2,3$), 
where the resultant $3\times 3$ blocks are\cite{catano2012neutrino,dias2012simple}
\begin{equation}
\label{eq:Neutrino-block-mass-matrices}
\begin{split}
m_{\nu} &=	\mathcal{M}_{\nu}^{\mathrm{T}} \left( \mathcal{M}_{\mathcal{N}} \right)^{-1} M_{\mathcal{N}} \left( \mathcal{M}_{\mathcal{N}}^{\mathrm{T}} \right)^{-1} \mathcal{M}_{\nu},	\\
M_{N} &\approx \mathcal{M}_{\mathcal{N}}-{M}_{\mathcal{N}},	\quad	
M_{\tilde{N}}\approx \mathcal{M}_{\mathcal{N}}+{M}_{\mathcal{N}}.
\end{split}
\end{equation}
The left-handed or active neutrinos $\nu_{L}^{\alpha}$ received tiny contributions from the right-handed or sterile species directly proportional to $M_{\mathcal{N}}$ and inversely to $\mathcal{M}_{\mathcal{N}}$, so as they get into the light neutrinos $\nu_{L}^{i}$ at thousandths of eV. On the contrary, the sterile species $\nu_{R}^{\alpha}$ and $\mathcal{N}_{R}^{\alpha}$ mixed them together into the heavy neutrinos $N_{R}^{i}$ and $\tilde{N}_{R}^{i}$ at units of TeV quasi-degenerated in ${M}_{N}$. 
                                                                                
Regarding one-loop corrections, the Ref. \cite{sierra2011importance} shows that neutrino mass matrix can be expressed by 
\begin{equation}
\mathbb{M}_{N} = \mathbb{M}_{N}^{\mathrm{tree}} + \mathbb{M}_{N}^{\mathrm{1-loop}},
\end{equation}
where $\mathbb{M}_{N}^{\mathrm{1-loop}}$ contains mass terms obtained from one-loop corrections
\begin{equation}
\mathbb{M}_{N}^{\mathrm{1-loop}} = 
\left(\begin{array}{c c}
\delta\boldsymbol{\mathcal{M}}_{L}	&	\delta\boldsymbol{\mathcal{M}}_{D}^{\mathrm{T}}	\\
\delta\boldsymbol{\mathcal{M}}_{D}	&	\delta\boldsymbol{\mathcal{M}}_{R}
\end{array}\right). 
\end{equation}
In this way, active neutrinos acquire masses through seesaw contributions at tree-level as well as through one-loop corrections 
\begin{equation}
m_{\nu} =	-\boldsymbol{\mathcal{M}}_{\nu}^{\mathrm{T}}\boldsymbol{\mathcal{M}}_{\mathcal{N}}^{-1}\boldsymbol{\mathcal{M}}_{\nu} + \delta\boldsymbol{\mathcal{M}}_{L}. 
\end{equation}
The one-loop mass $\delta\boldsymbol{\mathcal{M}}_{L}$ is obtained from radiative corrections on self-energy of left-handed neutrinos $\Sigma_{L}(p^{2}=0)$. It has contributions from $Z_{\mu}$, $G_{Z}$ and $h$, 
\begin{equation}
-i \Sigma(0) = - i\Sigma^{Z}(0) - i\Sigma^{G_{Z}}(0) - i\Sigma^{h}(0),
\end{equation}
and after regularizing the integrals involved in loop diagrams the correction turns out to be
\begin{eqnarray}
&&\delta\boldsymbol{\mathcal{M}}_{L} = 
\boldsymbol{\mathcal{M}}_{\nu}^{\mathrm{T}}\boldsymbol{\mathcal{M}}_{\mathcal{N}}^{-1} \\&&
\left\lbrace \frac{g^{2}}{64 \pi^{2} M_{W}^{2}}
	\left[ m_{h}^{2} 
	\ln\left(\frac{\boldsymbol{\mathcal{M}}_{\mathcal{N}}^{2}}{m_{h}^{2}}\right) + 
	3 m_{Z}^{2}
	\ln\left(\frac{\boldsymbol{\mathcal{M}}_{\mathcal{N}}^{2}}{m_{Z}^{2}}\right) \right]
\right\rbrace
\boldsymbol{\mathcal{M}}_{\nu}, \nonumber 
\end{eqnarray}
in which the heavy modes have been integrated out. Compared with tree-level expression, such a correction is suppressed by $\frac{1}{16\pi^{2}}\ln\left(\frac{\boldsymbol{\mathcal{M}}_{\mathcal{N}}^{2}}{m_{Z}^{2}}\right)$, so as it could contribute to the actual mass of light neutrinos. 

The model presented in this article has received some modification to forbid seesaw mechanisms at tree-level because one of the sterile neutrinos does not get mass and the mass matrix $\boldsymbol{\mathcal{M}}_{\mathcal{N}}$ does not have inverse. Consequently, it is mandatory to obtain masses through radiative corrections, in contrast with Ref. \cite{sierra2011importance} for normal SSM and \cite{dev2012minimal} for ISS, where seesaw is allowed at tree level and one-loop corrections are done in order to improve the predictability of the models. The section \ref{sect:Particle-content} introduces the scalar and leptonic content employed in the article as well as the neutral lepton Yukawa Lagrangian. Then, section \ref{sect:Radiative-masses} presents one-loop radiative corrections to get Majorana masses for light $\nu_{L}^{\alpha}$ and heavy $\nu_{R}^{\alpha}$ species. Finally, some conclusions are outlined in section \ref{sect:Conclusions}.

\section*{Scalar and leptonic content}
\label{sect:Particle-content}
\begin{table}[H]
\centering
\begin{tabular}{cccc}
Fermion	&	$X^{\pm}$	&	Bosons	&	$X^{\pm}$	\\	\hline	\hline
\multicolumn{2}{c}{Lepton doublets}	&	\multicolumn{2}{c}{Scalar doublets}	\\	\hline	\hline
$\small{\ell^{e}_{L}=\begin{pmatrix} \nu^{e} \\ e^{e} \end{pmatrix}_{L}}$
	&	$0^{+}$	&
$\small{\Phi_{1}=\begin{pmatrix}
\phi_{1}^{+} \\ \frac{h_{1}+v_{1}+i\eta_{1}}{\sqrt{2}}	
\end{pmatrix}}$	&	$\sfrac{+2}{3}^{+}$	\\
$\small{\ell^{\mu}_{L}=\begin{pmatrix} \nu^{\mu} \\ e^{\mu} \end{pmatrix}_{L}}$
	&	$0^{+}$	&
$\small{\Phi_{2}=\begin{pmatrix}
\phi_{2}^{+} \\ \frac{h_{2}+v_{2}+i\eta_{2}}{\sqrt{2}}	
\end{pmatrix}}$	&	$\sfrac{+1}{3}^{-}$	\\
$\small{\ell^{\tau}_{L}=\begin{pmatrix} \nu^{\tau} \\ e^{\tau} \end{pmatrix}_{L}}$
	&	$-1^{-}$	&
$\small{\Phi_{3}=\begin{pmatrix}
\phi_{3}^{+} \\ \frac{h_{3}+v_{3}+i\eta_{3}}{\sqrt{2}}	
\end{pmatrix}}$	&	$\sfrac{+1}{3}^{+}$	\\   \hline	\hline	\multicolumn{2}{c}{Lepton singlets}	&	\multicolumn{2}{c}{Scalar singlet}	\\	\hline	\hline
$\nu^{  e }_{R}$
	&	$\sfrac{+1}{3}^{+}$	&	
\multirow{3}{*}{
	\begin{tabular}{c}
		$\chi  =\frac{\xi_{\chi}  +v_{\chi}  +i\zeta_{\chi}}{\sqrt{2}}$\\$\sigma$
	\end{tabular}
}	&	
\multirow{3}{*}{
	\begin{tabular}{c}
		$\sfrac{-1}{3}^{+}$\\$\sfrac{-1}{3}^{-}$
	\end{tabular}
}	\\
$\nu^{\mu }_{R}$
	&	$\sfrac{+1}{3}^{+}$	\\
$\mathcal{N}^{\tau}_{R}$
	&	$\sfrac{+1}{3}^{+}$	\\	\hline	\hline
\multicolumn{2}{c}{Majorana fermions}	&	\multicolumn{2}{c}{Gauge bosons}	\\	\hline	\hline
$\mathcal{N}^{e}_{R}$	&	$0^{+}$	&
\small{$W^{\pm}_{\mu}$, $W^{0}_{\mu}$}	&	
\multirow{2}{*}{$0^{+}$}	\\
$\nu^{\tau}_{R}$
	&	$0^{+}$	&
$B_{\mu}$, $Z_{\mu}'$	\\	\hline	\hline
\end{tabular}
\caption{Bosonic and leptonic sector of the model with forbidden seesaw mechanism. The notation $X^{\pm}$ means the value of the $X$ charge and the $\mathbb{Z}_{2}$ parity. The VEVs have the hierarchy $v_{\chi}\ll v_{1}>v_{2}>v_{3}$\cite{mantilla2017nonuniversal}.}
\label{tab:Leptonic-sector}
\end{table}

The Ref. \cite{mantilla2017nonuniversal} introduces a new nonuniversal abelian extension to the SM where the fermion mass hierarchy is addressed. The majority of fermions acquire masses without radiative corrections, even light and heavy neutrinos through SSMs allowed by the Yukawa couplings of the model. However, the model presented here differs from the previous one in the neutral sector. The neutrinos $\nu_{R}^{\tau}\leftrightarrow N_{R}^{3}$ were interchanged so as the neutral lepton Yukawa Lagrangian turns out to be
\begin{equation}
\begin{split}
-\mathcal{L}_{N} &=
h_{3 \nu}^{e    e}\overline{\ell^{ e }_{L}}\tilde{\Phi}_{3}\nu^{ e  }_{R} + 
h_{3 \nu}^{e  \mu}\overline{\ell^{ e }_{L}}\tilde{\Phi}_{3}\nu^{\mu }_{R} + 
h_{3  N }^{ e\tau}\overline{\ell^{ e }_{L}}\tilde{\Phi}_{3}\mathcal{N}^{\tau}_{R} \\ &+
h_{3 \nu}^{\mu  e}\overline{\ell^{\mu}_{L}}\tilde{\Phi}_{3}\nu^{e}_{R} +
h_{3 \nu}^{\mu\mu}\overline{\ell^{\mu}_{L}}\tilde{\Phi}_{3}\nu^{\mu}_{R} + 
h_{3  N }^{\mu  \tau}\overline{\ell^{\mu}_{L}}\tilde{\Phi}_{3}\mathcal{N}^{\tau}_{R} \\ &+
g_{\chi \Scale[0.6]{N}}^{e e} 
\Scale[0.9]{\overline{\nu_{R}^{eC}}} 
\tilde\chi \Scale[0.9]{\mathcal{N}_{R}^{e}} +
g_{\chi \Scale[0.6]{N}}^{\mu e} 
\Scale[0.9]{\overline{\nu_{R}^{\mu C}}} 
\tilde\chi \Scale[0.9]{\mathcal{N}_{R}^{e}} +
g_{\chi \Scale[0.6]{N}}^{\tau e} 
\Scale[0.9]{\overline{\mathcal{N}_{R}^{\tau C}}} 
\tilde\chi \Scale[0.9]{\mathcal{N}_{R}^{e}} \\ &+
g_{\chi \Scale[0.6]{N}}^{e \tau} 
\Scale[0.9]{\overline{\nu_{R}^{eC}}} 
\tilde\sigma \Scale[0.9]{\nu^{\tau}_{R}} + 
g_{\chi \Scale[0.6]{N}}^{\mu  \tau} 
\Scale[0.9]{\overline{\nu_{R}^{\mu C}}} 
\tilde\chi \Scale[0.9]{\nu^{\tau}_{R}} + 
g_{\chi \Scale[0.6]{N}}^{\tau\tau} 
\Scale[0.9]{\overline{\mathcal{N}_{R}^{\tau C}}} 
\tilde\chi \Scale[0.9]{\nu^{\tau}_{R}}\\ & + 
\Scale[1.0]{\frac{M^{ee}_{N}}{2\sqrt{2}}}
\Scale[0.9]{\overline{\mathcal{N}_{R}^{e C}}}
\Scale[0.9]{\mathcal{N}_{R}^{e}} + \Scale[1.0]{\frac{M^{e\tau}_{N}}{2\sqrt{2}}}
\Scale[0.9]{\overline{\mathcal{N}_{R}^{e C}}}
\Scale[0.9]{\nu^{\tau}_{R}} + 
\Scale[1.0]{\frac{M^{\tau\tau}_{\nu}}{2\sqrt{2}}}
\Scale[0.9]{\overline{\nu_{R}^{\tau C}}}
\Scale[0.9]{\nu^{\tau}_{R}} + 
\mathrm{h.c.},
\end{split}
\label{eq:Neutrino-Lagrangian}
\end{equation}
where $\widetilde{\Phi}=i\sigma_2 \Phi^*$ are the scalar doublet conjugates and the Majorana mass is denoted by $M^{ij}_{\mathcal{N}}$. The neutral sector of the model involves Dirac and Majorana masses in its Yukawa Lagrangian. They can be described in the flavor and mass bases which are, respectively,
\begin{equation}
\begin{split}
\mathbf{N}_{L}&=(\nu^{e,\mu,\tau}_{L},{\nu^{e,\mu,\tau}_{R}}^{C},{\mathcal{N}^{e,\tau}_{R}}^{C}),\\
\mathbf{n}_{L}&=(\nu^{1,2,3}_{L},{N^{1,2,3}_{R}}^{C},\left.\mathcal{N}^{1,3}_{R}\right.^{C}).
\end{split}
\end{equation}
After the SSB the mass terms of the neutral sector in the flavor basis are
\begin{equation}
-\mathcal{L}_{N} = \frac{1}{2} \overline{\mathbf{N}_{L}^{C}} \mathbb{M}_{N}\mathbf{N}_{L},
\end{equation}
where the mass matrix has the following block structure
\begin{equation}
\mathbb{M}_{N} = 
\left(\begin{array}{c c c}
0	&	\boldsymbol{\mathcal{M}_{\nu}}^{\mathrm{T}}	\\
\boldsymbol{\mathcal{M}}_{\nu}	&	\boldsymbol{\mathcal{M}}_{\mathcal{N}}
\end{array}\right),
\end{equation}
where the block $\boldsymbol{\mathcal{M}}_{\nu}$ is
\begin{equation}
\label{eq:m_nu_original_parameters}
\mathcal{M}_{\nu}^{\mathrm{T}} = \frac{v_{3}}{\sqrt{2}}
\left(
\begin{array}{ccc|ccc}
h_{3 \nu}^{e e}	&	h_{3 \nu}^{e \mu}	&0 & 0	&	h_{3  N }^{ e \tau }\\
h_{3 \nu}^{\mu e}&	h_{3 \nu}^{\mu \mu}	&0 &   0	&	h_{3  N }^{\mu \tau } \\
0	&	0	&0&   0&	0
\end{array}
\right),
\end{equation}
while the block $\boldsymbol{\mathcal{M}}_{\mathcal{N}}$ turns out to be
\begin{equation}
\mathcal{M}_{\mathcal{N}} = \frac{v_{\chi}}{\sqrt{2}}
\left(
\begin{array}{ccc|ccc}
0	&	0	&	g_{\Scale[0.7]{\chi} \Scale[0.6]{N}}^{e \tau}	&
g_{\chi \Scale[0.6]{N}}^{e e}   &	0	\\
0	&	0	&	g_{\Scale[0.7]{\chi} \Scale[0.6]{N}}^{\mu \tau}	&
g_{\chi \Scale[0.6]{N}}^{\mu e} &   0	\\
g_{\Scale[0.7]{\chi} \Scale[0.6]{N}}^{e \tau}	&	
g_{\Scale[0.7]{\chi} \Scale[0.6]{N}}^{\mu \tau}	&	
\Scale[0.9]{\tilde{M}^{\tau\tau}_{\nu}}	&	
\Scale[0.9]{\tilde{M}^{e\tau}_{N}}  &   
g_{\chi \Scale[0.6]{N}}^{\tau\tau} 	\\
\hline 
g_{\chi \Scale[0.6]{N}}^{e e}	&	
g_{\chi \Scale[0.6]{N}}^{\mu e}	&	
\Scale[0.9]{\tilde{M}^{e\tau}_{N}}	&
\Scale[0.9]{\tilde{M}^{ee}_{N}}	    &   
g_{\chi \Scale[0.6]{N}}^{\tau e}
 	\\
0	&	
0	&	
g_{\chi \Scale[0.6]{N}}^{\tau\tau} 	&
g_{\chi \Scale[0.6]{N}}^{\tau e} 	&	
0
\end{array}
\right)
\end{equation}
where $\tilde{M}_{N}^{ij}=M_{N}^{ij}/v_{\chi}$. The determinant of $\mathcal{M}_{\mathcal{N}}$ is null. Moreover, the active neutrino $\nu^{\tau}_{L}$ is disconnected from the rest of the neutral fermion system, leaving it massless and constraining the model to inverse ordering (IO)\cite{esteban2017updated,nufit} with $\nu^{\tau}_{L}=\nu^{3}_{L}$. It is important to note that in the Yukawa Lagrangian \eqref{eq:Neutrino-Lagrangian} the leptonic doublet $\ell_{L}^{\tau}$ does not appear due to its $X$-charge which was assigned in order to cancel chiral anomalies. Consequently, it cannot acquire mass neither by see-saw nor radiative corrections. 

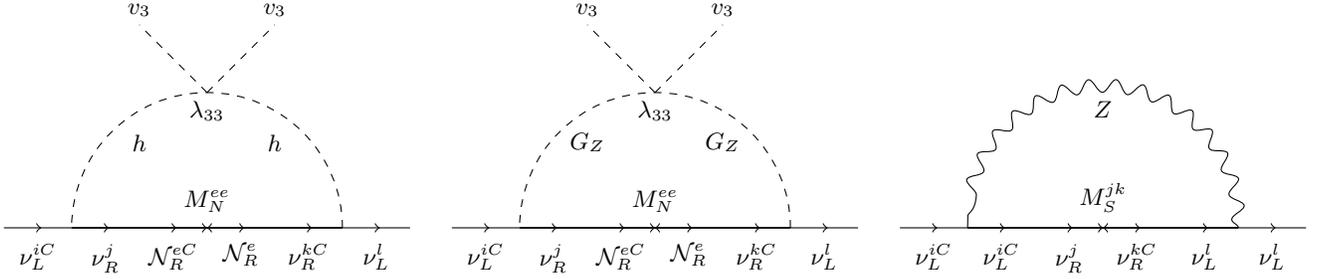
\begin{figure*}[h]
\centering


\begin{tikzpicture}[node distance=0.9cm and 0.9cm]
\coordinate[        ] (v0);
\coordinate[        right=of v0] (v1);
\coordinate[        right=of v1] (v2);
\coordinate[        right=of v2] (v4);
\coordinate[        right=of v2] (v3);
\coordinate[        right=of v4] (v6);
\coordinate[        right=of v4] (v5);
\coordinate[        right=of v6] (v7);
\coordinate[        right=of v7] (v8);
\coordinate[        above=of v2,label=$h$] (vHL);
\coordinate[        above=of v6,label=$h$] (vHR);
\coordinate[        above=of v4] (vXa);
\coordinate[        above=of vXa] (vXb);
\coordinate[        right=of vXb] (vR);
\coordinate[        left =of vXb] (vL);
\coordinate[        above=of vR,label=above:$v_{3}$] (vYR);
\coordinate[        above=of vL,label=above:$v_{3}$] (vYL);
\coordinate[        above=of vXa,label=below:$\lambda_{33}$] (vZ);
\draw[fermion] (v0) -- (v1) node[midway,below=0.1cm] {$\nu^{i C}_{L}$};
\draw[fermion] (v1) -- (v2) node[midway,below=0.1cm] {$\nu^{j}_{R}$};
\draw[fermion] (v2) -- (v4) node[midway,below=0.1cm] {$\mathcal{N}_{R}^{1 C}$};
\draw[fermion] (v4) -- (v6) node[midway,below=0.1cm] {$\mathcal{N}_{R}^{1}$};
\draw[fermion] (v6) -- (v7) node[midway,below=0.1cm] {$\nu_{R}^{k C}$};
\draw[fermion] (v7) -- (v8) node[midway,below=0.1cm] {$\nu^{l}_{L}$};

\draw[fermionA] (v1) -- (v7) node[midway,below=0.1cm] {};
\draw[fermionA] (v7) -- (v1) node[midway,above=0.1cm] {$M^{ee}_{N}$};



\draw[dashed] (vXb) -- (vYR); 
\draw[dashed] (vXb) -- (vYL); 
\semiloopBelow[dashed]{v1}{v7}{0};
\end{tikzpicture}
$\quad$
\begin{tikzpicture}[node distance=0.9cm and 0.9cm]
\coordinate[        ] (v0);
\coordinate[        right=of v0] (v1);
\coordinate[        right=of v1] (v2);
\coordinate[        right=of v2] (v4);
\coordinate[        right=of v2] (v3);
\coordinate[        right=of v4] (v5);
\coordinate[        right=of v4] (v6);
\coordinate[        right=of v6] (v7);
\coordinate[        right=of v7] (v8);
\coordinate[        above=of v2,label=$G_{Z}$] (vHL);
\coordinate[        above=of v6,label=$G_{Z}$] (vHR);
\coordinate[        above=of v4] (vXa);
\coordinate[        above=of vXa] (vXb);
\coordinate[        right=of vXb] (vR);
\coordinate[        left =of vXb] (vL);
\coordinate[        above=of vR,label=above:$v_{3}$] (vYR);
\coordinate[        above=of vL,label=above:$v_{3}$] (vYL);
\coordinate[        above=of vXa,label=below:$\lambda_{33}$] (vZ);
\draw[fermion] (v0) -- (v1) node[midway,below=0.1cm] {$\nu^{i C}_{L}$};
\draw[fermion] (v1) -- (v2) node[midway,below=0.1cm] {$\nu^{j}_{R}$};
\draw[fermion] (v2) -- (v4) node[midway,below=0.1cm] {$\mathcal{N}_{R}^{1 C}$};
\draw[fermion] (v4) -- (v6) node[midway,below=0.1cm] {$\mathcal{N}_{R}^{1}$};
\draw[fermion] (v6) -- (v7) node[midway,below=0.1cm] {$\nu_{R}^{k C}$};
\draw[fermion] (v7) -- (v8) node[midway,below=0.1cm] {$\nu^{l}_{L}$};

\draw[fermionA] (v1) -- (v7) node[midway,below=0.1cm] {};
\draw[fermionA] (v7) -- (v1) node[midway,above=0.1cm] {$M^{ee}_{N}$};
\draw[dashed] (vXb) -- (vYR); 
\draw[dashed] (vXb) -- (vYL); 
\semiloopBelow[dashed]{v1}{v7}{0};
\end{tikzpicture}
$\quad$
\begin{tikzpicture}[node distance=0.9cm and 0.9cm]
\coordinate[        ] (v0);
\coordinate[        right=of v0] (v1);
\coordinate[        right=of v1] (v2);
\coordinate[        right=of v2] (v4);
\coordinate[        right=of v4] (v6);
\coordinate[        right=of v6] (v7);
\coordinate[        right=of v7] (v8);
\coordinate[        above=of v5] (vXa);
\coordinate[        above=of vXa] (vXb);
\coordinate[        left =of vXb] (vL);
\draw[fermion] (v0) -- (v1) node[midway,below=0.1cm] {$\nu^{i C}_{L}$};
\draw[fermion] (v1) -- (v2) node[midway,below=0.1cm] {$\nu^{i C}_{L}$};
\draw[fermion] (v2) -- (v4) node[midway,below=0.1cm] {$\nu_{R}^{j}$};
\draw[fermion] (v4) -- (v6) node[midway,below=0.1cm] {$\nu_{R}^{k C}$};
\draw[fermion] (v6) -- (v7) node[midway,below=0.1cm] {$\nu^{l}_{L}$};
\draw[fermion] (v7) -- (v8) node[midway,below=0.1cm] {$\nu^{l}_{L}$};

\draw[fermionA] (v1) -- (v7) node[midway,below=0.1cm] {};
\draw[fermionA] (v7) -- (v1) node[midway,above=0.1cm] {$M^{jk}_{S}$};
\semiloopBelow[photon]{v1}{v7}{0}[$Z$];
\end{tikzpicture}


\caption{Radiative corrections at one-loop in the self-energy of $\nu_{L}$ where $i,l=e,\mu$ and $j,k=e,\mu,\tau$.}
\label{fig:Loop-nL}
\end{figure*}

\section*{Radiative masses}
\label{sect:Radiative-masses}
The impossibility to perform the SSM in the neutral sector of the model implies massless
active neutrinos at tree level. However, there exist one-loop diagrams which are suitable to get Majorana masses for active neutrinos involving right-handed neutrinos and the lepton number violation Majorana mass terms $M_{N}$. The next sections show how the model accounts for $\nu_
{L}^{\alpha}$ masses as well as right-handed neutrinos.

\subsection*{Radiative mass for active neutrinos $\nu_{L}$}
Since there is one right-handed neutrino massless at tree-level, the SSM is forbidden and the masses of light neutrinos have to be obtained through radiative corrections at one-loop level. The main contributions to the self-energy of $\nu_{L}^{\alpha}$ come from SM fields: the gauge boson $Z_{\mu}$, its Goldstone boson $G_{Z}$ and the Higgs boson $h$, so as\cite{grimus2002one,sierra2011importance,dev2012minimal}
\begin{equation}
-i \Sigma(p^{2}=0) = - i\Sigma^{Z}(0) - i\Sigma^{G_{Z}}(0) - i\Sigma^{h}(0). 
\end{equation}
The contributions from non-SM fields are suppressed because of the hierarchy among $v$, the electroweak VEV, and $v_{\chi}$ at TeV scale. For instance, in the scalar sector the flavor basis $\mathbf{h}=(h_{1},h_{2},h_{3},\xi_{\chi})$ is connected with the mass basis $\mathbf{H}=(h,H_{1},H_{2},\mathcal{H})$ through $\mathbf{H}=R_{\mathrm{even}}\mathbf{h}$ (the Higgs potential as well as scalar masses and mixings are explained in detail in Ref. \cite{mantilla2017nonuniversal}). After the SSB, there are four CP-even scalar bosons: the lightest corresponding with the SM Higgs $v$, two intermediate scalars $H_{1}$ and $H_{2}$ at TeV scale and the heaviest $\mathcal{H}$ associated with the SSB of $\mathrm{U(1)}_{X}$. The mixing matrix $R_{\mathrm{even}}$ can be expressed as 
\begin{align}
R_{\mathrm{even}}&=
\begin{pmatrix}
	1	&	s_{12}^h	&	s_{13}^h	&	s_{14}^{h} \\
 -s_{12}^h & 1 & 0	&	s_{24}^{h} \\
 -s_{13}^h & 0 & 1	&	s_{34}^{h} \\
 -s_{14}^h	&	-s_{24}^h	&	-s_{34}^h	&	1
\end{pmatrix}
\end{align}
where the mixing angles are given by
\begin{equation}
s_{i4}^{h}\approx \frac{v_{i}}{v_{\chi}},	\qquad 
s_{12}^{h} \approx \frac{v_{2}}{v_{1}}, \qquad 
s_{13}^{h}\approx \frac{v_{3}}{v_{1}}.
\end{equation}
The most suppressed mixings involve $s_{i4}^{h}$ because $v_{i}\ll v_{\chi}$, followed by $s_{13}^{h}$ and $s_{12}^{h}$. Moreover, since light neutrinos couple with $h_{3}$ (look at the two first rows in Eq. \eqref{eq:Neutrino-Lagrangian}), which is expressed in the eigenstate basis as the superposition of $h$ and the heavier scalars $H_{2}$ and $\mathcal{H}$, then
\begin{equation}
h_{3} \approx H_{2} + s_{13}^{h} h - s_{34} \mathcal{H},
\end{equation}
the less suppressed contribution to light neutrinos self-energy comes from the lightest scalar, $h$, with the mixing angle $s_{13}^{h}$. 

Now, regarding the diagrams in Fig. \ref{fig:Loop-nL}, the fermionic propagator inside the loop is composed by four Yukawa coupling constants, two mass insertions between $\nu_{R}$ and $\mathcal{N}_{R}$ and the Majorana mass $M_{N}^{ee}$ which violates lepton number conservation and connects $\nu_{L}$ with $\nu_{L}^{C}$\cite{dev2012minimal,sierra2011importance}. It is possible to associate an effective mass $M_{S}$ to the fermionic propagator 
\begin{equation}
M_{S}^{jk}= g_{\chi \Scale[0.6]{N}}^{k e} g_{\chi \Scale[0.6]{N}}^{j e*}\frac{v_{\chi}^{2}}{2 M_{N}^{ee}}.
\end{equation}

The algebraic expression of the first diagram in Fig. \ref{fig:Loop-nL}, the self-energy involving the Higgs boson, is
\begin{equation}
\begin{split}
& P_{R} i [\Sigma(\cancel{p})^{h}]^{li}P_{R} = 
\lambda_{33}v_{3}^{2}
\frac{h_{3 \nu}^{lk}s_{13}^{h}}{\sqrt{2}} P_{R}	\\&	
\int \frac{dk}{(2\pi)^{4}}
\frac{\cancel{p}-\cancel{k}+M_{S}^{kj}}{(p-k)^{2}-(M_{S}^{kj})^{2}}
\frac{1}{(k^{2}-m_{h}^{2})^{2}} P_{R} 
\frac{h_{3 \nu}^{lk*}s_{13}^{h}}{\sqrt{2}}
\end{split}
\end{equation}
where $\lambda_{33}$ is the quartic coupling constant of $(\Phi_{3}^{\dagger}\Phi_{3})^{2}$ in the Higgs potential which can be approximated by
\begin{equation}
\lambda_{33} \approx \frac{m_{h}^{2}}{4 M_{W}^{2}}
\end{equation}
On the other hand, from the expressions of the second and third diagrams in Fig. \ref{fig:Loop-nL}, the self-energy involving $G_{Z}$ and $Z_{\mu}$ are given by
\begin{equation}
\begin{split}
& P_{R} i [\Sigma(\cancel{p})^{Z}]^{li}P_{R} = 
\frac{g^{2}}{4c_{W}^{2}}
\frac{h_{3 \nu}^{lk}v_{3}^{h}}{\sqrt{2}} P_{R}	\\&	
\gamma_{\mu}
\int \frac{dk}{(2\pi)^{4}}
\frac{\cancel{p}-\cancel{k}}{(p-k)^{2}}
\frac{\cancel{p}-\cancel{k}+M_{S}^{kj}}{(p-k)^{2}-(M_{S}^{kj})^{2}}	\\&	
\frac{\cancel{p}-\cancel{k}}{(p-k)^{2}} P_{R} 
\frac{g^{\mu\nu}}{(k^{2}-m_{Z}^{2})^{2}}
\gamma_{\nu}
\frac{h_{3 \nu}^{lk*}v_{3}^{h}}{\sqrt{2}}.
\end{split}
\end{equation}

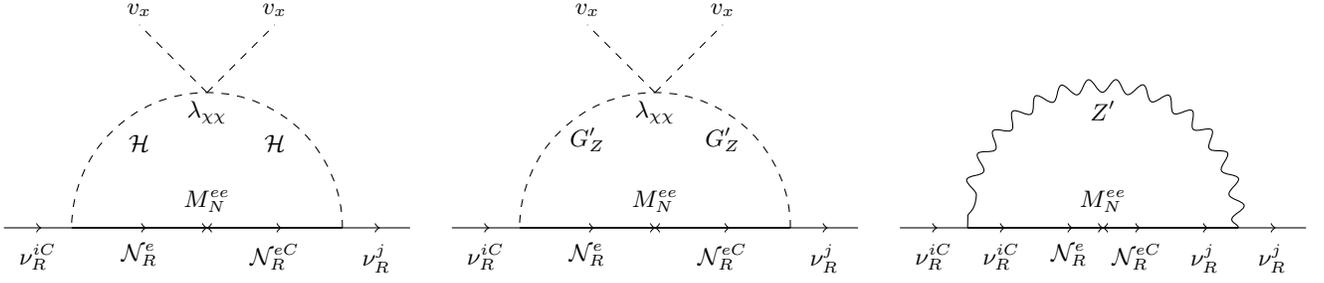
\begin{figure*}[h]
\centering

\begin{tikzpicture}[node distance=0.9cm and 0.9cm]
\coordinate[        ] (v0);
\coordinate[        right=of v0] (v1);
\coordinate[        right=of v1] (v2);
\coordinate[        right=of v2] (v4);
\coordinate[        right=of v2] (v3);
\coordinate[        right=of v4] (v6);
\coordinate[        right=of v4] (v5);
\coordinate[        right=of v6] (v7);
\coordinate[        right=of v7] (v8);
\coordinate[        above=of v2,label=$\mathcal{H}$] (vHL);
\coordinate[        above=of v6,label=$\mathcal{H}$] (vHR);
\coordinate[        above=of v4] (vXa);
\coordinate[        above=of vXa] (vXb);
\coordinate[        right=of vXb] (vR);
\coordinate[        left =of vXb] (vL);
\coordinate[        above=of vR,label=above:$v_{x}$] (vYR);
\coordinate[        above=of vL,label=above:$v_{x}$] (vYL);
\coordinate[        above=of vXa,label=below:$\lambda_{\chi\chi}$] (vZ);
\draw[fermion] (v0) -- (v1) node[midway,below=0.1cm] {$\nu^{i C}_{R}$};
\draw[fermion] (v1) -- (v4) node[midway,below=0.1cm] {$\mathcal{N}_{R}^{1}$};
\draw[fermion] (v4) -- (v7) node[midway,below=0.1cm] {$\mathcal{N}_{R}^{1 C}$};
\draw[fermion] (v7) -- (v8) node[midway,below=0.1cm] {$\nu^{j}_{R}$};

\draw[fermionA] (v1) -- (v7) node[midway,below=0.1cm] {};
\draw[fermionA] (v7) -- (v1) node[midway,above=0.1cm] {$M^{ee}_{N}$};
\draw[dashed] (vXb) -- (vYR); 
\draw[dashed] (vXb) -- (vYL); 
\semiloopBelow[dashed]{v1}{v7}{0};
\end{tikzpicture}
$\quad$
\begin{tikzpicture}[node distance=0.9cm and 0.9cm]
\coordinate[        ] (v0);
\coordinate[        right=of v0] (v1);
\coordinate[        right=of v1] (v2);
\coordinate[        right=of v2] (v4);
\coordinate[        right=of v2] (v3);
\coordinate[        right=of v4] (v5);
\coordinate[        right=of v4] (v6);
\coordinate[        right=of v6] (v7);
\coordinate[        right=of v7] (v8);
\coordinate[        above=of v2,label=$G_{Z}'$] (vHL);
\coordinate[        above=of v6,label=$G_{Z}'$] (vHR);
\coordinate[        above=of v4] (vXa);
\coordinate[        above=of vXa] (vXb);
\coordinate[        right=of vXb] (vR);
\coordinate[        left =of vXb] (vL);
\coordinate[        above=of vR,label=above:$v_{x}$] (vYR);
\coordinate[        above=of vL,label=above:$v_{x}$] (vYL);
\coordinate[        above=of vXa,label=below:$\lambda_{\chi\chi}$] (vZ);
\draw[fermion] (v0) -- (v1) node[midway,below=0.1cm] {$\nu^{i C}_{R}$};
\draw[fermion] (v1) -- (v4) node[midway,below=0.1cm] {$\mathcal{N}_{R}^{1}$};
\draw[fermion] (v4) -- (v7) node[midway,below=0.1cm] {$\mathcal{N}_{R}^{1 C}$};
\draw[fermion] (v7) -- (v8) node[midway,below=0.1cm] {$\nu^{j}_{R}$};

\draw[fermionA] (v1) -- (v7) node[midway,below=0.1cm] {};
\draw[fermionA] (v7) -- (v1) node[midway,above=0.1cm] {$M^{ee}_{N}$};
\draw[dashed] (vXb) -- (vYR); 
\draw[dashed] (vXb) -- (vYL); 
\semiloopBelow[dashed]{v1}{v7}{0};
\end{tikzpicture}
$\quad$
\begin{tikzpicture}[node distance=0.9cm and 0.9cm]
\coordinate[        ] (v0);
\coordinate[        right=of v0] (v1);
\coordinate[        right=of v1] (v2);
\coordinate[        right=of v2] (v4);
\coordinate[        right=of v4] (v6);
\coordinate[        right=of v6] (v7);
\coordinate[        right=of v7] (v8);
\coordinate[        above=of v5] (vXa);
\coordinate[        above=of vXa] (vXb);
\coordinate[        left =of vXb] (vL);
\draw[fermion] (v0) -- (v1) node[midway,below=0.1cm] {$\nu^{i C}_{R}$};
\draw[fermion] (v1) -- (v2) node[midway,below=0.1cm] {$\nu^{i C}_{R}$};
\draw[fermion] (v2) -- (v4) node[midway,below=0.1cm] {$\mathcal{N}_{R}^{1}$};
\draw[fermion] (v4) -- (v6) node[midway,below=0.1cm] {$\mathcal{N}_{R}^{1 C}$};
\draw[fermion] (v6) -- (v7) node[midway,below=0.1cm] {$\nu^{j}_{R}$};
\draw[fermion] (v7) -- (v8) node[midway,below=0.1cm] {$\nu^{j}_{R}$};

\draw[fermionA] (v1) -- (v7) node[midway,below=0.1cm] {};
\draw[fermionA] (v7) -- (v1) node[midway,above=0.1cm] {$M^{ee}_{N}$};
\semiloopBelow[photon]{v1}{v7}{0}[$Z'$];
\end{tikzpicture}

\caption{Radiative corrections at one-loop in the self-energy of $\nu_{R}$ where $i,j=e,\mu,\tau$.}
\label{fig:Loop-nR}
\end{figure*}

After integrating both contributions, the resulting mass for light neutrinos is
\begin{equation}
\begin{split}
M_{\nu}^{li} &= \frac{\alpha_{W}}{64 \pi	M_{W}^{2}}
\left( h_{3 \nu} g_{\chi \Scale[0.6]{N}} g_{\chi \Scale[0.6]{N}}^{*} h_{3 \nu}^{*} \right)^{li}
\frac{v_{3}^{2}v_{\chi}^{2}}{M^{ee}_{N}} (s_{13}^{h})^{2}	\\&	
\left\lbrace 
\frac{m_{h}^{2}}{M_{S}^{2}-m_{h}^{2}}
\ln \frac{M_{S}^{2}}{m_{h}^{2}} + 3
\frac{m_{Z}^{2}}{M_{S}^{2}-m_{Z}^{2}}
\ln \frac{M_{S}^{2}}{m_{Z}^{2}} \right\rbrace.
\end{split}
\end{equation}
The expression can be simplified by defining $\tilde{M}_{S}=v_{\chi}^{2}/M_{N}^{ee}$ and $s_{\gamma}=v_{3}/v$, and by replacing $M_{W}=gv/2$ so as
\begin{equation}
\frac{\alpha_{W}v_{3}^{2}}{64 \pi	M_{W}^{2}} = 
\frac{1}{64\pi^{2}} \frac{v_{3}^{2}}{v_{\chi}^{2}} = 
\frac{1}{64\pi^{2}} \sin^{2}_{\gamma}
\end{equation}
and the mass matrix is
\begin{equation}
\begin{split}
M_{\nu}^{li} &= \frac{\sin^{2}_{\gamma}(s_{13}^{h})^{2}}{64\pi^{2}} \tilde{M}_{S}
\left( h_{3 \nu} g_{\chi \Scale[0.6]{N}} g_{\chi \Scale[0.6]{N}}^{*} h_{3 \nu}^{*} \right)^{li}	\\&	
\left\lbrace 
\frac{m_{h}^{2}}{M_{S}^{2}-m_{h}^{2}}
\ln \frac{M_{S}^{2}}{m_{h}^{2}} + 3
\frac{m_{Z}^{2}}{M_{S}^{2}-m_{Z}^{2}}
\ln \frac{M_{S}^{2}}{m_{Z}^{2}} \right\rbrace.
\end{split}
\end{equation}

Some interesting limits on $M_{\nu}^{li}$ can be drafted depending on the hierarchy between $M_{N}^{1}$ and $v_{\chi}$. The fist one assumes $g_{\chi \Scale[0.6]{N}}v_{\chi}\ll M_{N}^{ee}$, so the mass matrix turns out to be
\begin{equation}
\begin{split}
M_{\nu}^{li} &= \frac{s^{2}_{\gamma}(s^{h}_{13})^{2}}{16\pi^{2}}
\frac{\left( h_{3 \nu} g_{\chi \Scale[0.6]{N}} g_{\chi \Scale[0.6]{N}}^{*} h_{3 \nu}^{*} \right)^{li}}
{g_{\chi \Scale[0.6]{N}}^{k e} g_{\chi \Scale[0.6]{N}}^{j e*}g_{\chi \Scale[0.6]{N}}^{j e} g_{\chi \Scale[0.6]{N}}^{k e*}}	\\&	
\left\lbrace 
\frac{m_{h}^{2}+3 m_{Z}^{2}}{\tilde{M_{S}}}
\ln \frac{\tilde{M}_{S}^{2}}{m_{h}^{2}}
\right\rbrace
\end{split}
\end{equation}
where $\tilde{M}_{S}$ can be large as it is required to obtain light neutrinos. On the other hand, when $M_{N}^{ee} \ll g_{\chi \Scale[0.6]{N}}v_{\chi}\ll m_{h}$
\begin{equation}
\begin{split}
M_{\nu}^{li} = \frac{s^{2}_{\gamma}(s^{h}_{13})^{2}}{ 8\pi^{2}}
\left( h_{3 \nu} g_{\chi \Scale[0.6]{N}} g_{\chi \Scale[0.6]{N}}^{*} h_{3 \nu}^{*} \right)^{li}
\tilde{M}_{S}\ln \frac{m_{h}^{2}}{\tilde{M}_{S}^{2}}
\end{split}
\end{equation}
where, in this case, $\tilde{M}_{S}$ gets small in consistency with light neutrinos. 

\subsection*{Radiative mass for the massless $\nu_{R}$}
An analogous calculation is done on right-handed sector by doing some replacements on radiative correction and similar expressions are obtained, where the SM scalar and gauge fields $h$, $G_{Z}$ and $Z_{\mu}$ have been replaced by the new $\mathcal{H}$, $G_{Z}'$ and $Z_{\mu}'$, respectively. The contributions from the SM Higgs boson $h$ get suppressed because of the small mixing angles $v_{i}/v_{\chi}$ with $\mathcal{H}$, which receives major contributions from $\xi_{\chi}$
\begin{equation}
\mathcal{H} = \xi_{\chi} - \sum_{i=1}^{3}\frac{v_{i}h_{i}}{v_{\chi}}. 
\end{equation}
Similarly, $Z_{\mu}$ does not contribute importantly to the self-energy of $\nu_{R}$ because of the mixing $v^{2}/v_{\chi}^{2}$ with $Z_{\mu}'$, so SM bosons can be neglected in first approximation. 

Since the diagrams of Figs. \ref{fig:Loop-nL} and \ref{fig:Loop-nR} are similar under the replacements $(h,G_{Z},Z_{\mu}) \rightarrow (\mathcal{H},G_{Z}',Z_{\mu}')$ with the suited coupling constants, the resulting mass matrix for right-handed neutrinos is
\begin{equation}
\begin{split}
M_{R}^{ki} &= \frac{g_{X}}{64 \pi^{2} {M_{Z}'}^{2}}
 g_{\chi \Scale[0.6]{N}}^{kj} g_{\chi \Scale[0.6]{N}}^{ji*}
\frac{v_{\chi}^{2}}{M^{ee}_{N}} \\&	
\left\lbrace 
\frac{m_{h}^{2}}{M_{S}^{2}-M_{\mathcal{H}}^{2}}
\ln \frac{M_{S}^{2}}{M_{\mathcal{H}}^{2}} + 3
\frac{m_{Z}^{2}}{M_{S}^{2}-M_{Z'}^{2}}
\ln \frac{M_{S}^{2}}{M_{Z'}^{2}} \right\rbrace
\end{split}
\end{equation}
Similarly with the left-handed neutrino sector, some limits can be drafted. When $M_{\mathcal{H},Z'}\approx v_{\chi} \ll M_{N}^{ee}$
\begin{equation}
\begin{split}
M_{R}^{ki} &= \frac{3}{16 \pi^{2}}
 g_{\chi \Scale[0.6]{N}}^{kj} g_{\chi \Scale[0.6]{N}}^{ji*}
\frac{M_{Z'}^{4}}{(M_{N}^{ee})^{3}}\ln\left( \frac{(M_{N}^{ee})^{2}}{M_{Z'}^{2}} \right)
\end{split}
\end{equation}
where the mass of the non-SM gauge boson $Z_{\mu}'$, given by $M_{Z'}\approx g_{X}v_{\chi}/3$, has been replaced\cite{mantilla2017nonuniversal}. This case yields the massless $\nu_{R}^{2}$ as a light right-handed sterile neutrino. On the other hand, when $M_{\mathcal{H},Z'}\gg M_{N}^{ee}$
\begin{equation}
\begin{split}
M_{R}^{ki} &= \frac{3}{16 \pi^{2}}
 g_{\chi \Scale[0.6]{N}}^{kj} g_{\chi \Scale[0.6]{N}}^{ji*}
\frac{M_{Z'}^{2}}{(M_{N}^{ee})}\ln\left( \frac{M_{Z'}^{2}}{(M_{N}^{ee})^{2}} \right)
\end{split}
\end{equation}
and the former massless $\nu_{R}^{2}$ would get mass near to the TeV scale. 

One interesting feature in one-loop mass expressions is the presence of terms $\frac{m_{h}^{2}}{(M_{S})}$ or $\frac{M_{Z'}^{2}}{(M_{N}^{ee})}$, suggesting the existence of a seesaw-like mechanism inside self-energy loops. Actually, those factors determine the mass scale of neutrinos which have acquired masses through radiative corrections. 

\section*{Discussion and Conclusions}
\label{sect:Conclusions}
The evidence of neutrino oscillations, as well as the upper limits on absolute neutrino masses have yielded stringent constraints in models which address neutrino physics. The most preferred framework to understand neutrino masses is the seesaw mechanism, whose main hypothesis is the existence of sterile right-handed neutrinos as Majorana fermions, which have lepton number violating masses at a higher scale so as left-handed neutrinos turns out to be Majorana fermions with masses at eV scale. Moreover, the introduction of a second set of right-handed neutrinos, together with the existence of a second VEV at TeV allows lower Majorana mass scale to accessible experiment energies. However, there exist the scenario in which the Majorana mass matrix does not have inverse matrix, yielding massless right-handed neutrinos as well as the impossibility to obtain neutrino masses at tree level. In this way, radiative corrections comprise the principal mechanism to get light massive neutrinos. 

The model presented here implies such a scenario where radiative corrections are the main mechanism to get masses. There are two kinds of loops, the self-energies of $\nu_{L}$ involving the SM bosons $Z_{\mu}$, $G_{Z}$ and $h$ together with right-handed neutrinos $\nu_{R}$ and $\mathcal{N}_{R}$, and the self-energy of $\nu_{R}^{2}$ involving the non-SM bosons $Z_{\mu}'$, $G_{Z}'$ and $\mathcal{H}$ with the Majorana fermions $\mathcal{N}_{R}$. Furthermore, depending on the hierarchy between the VEV $v_{\chi}$ at TEV scale and the Majorana mass $M_{N}^{ee}$ which would be at a higher or lower scale. Such limits were obtained in section \ref{sect:Radiative-masses} and are outlined in Tab. \ref{tab:Special-cases}. 

Consequently, the model shows how a scenario in which Majorana mass matrix does not have inverse implies massless neutrinos at tree level and massive at one-loop level through radiative corrections involving heavy modes and bosons, in contrast with the majority of models where neutrinos get masses at tree level by seesaw mechanisms. 

\begin{table}[h]
\centering
\begin{tabular}{ccc}
	&	$v_{\chi}\gg M_{N}^{ee}$	&	$v_{\chi}\ll M_{N}^{ee}$	\\\hline\hline
$\nu_{L}$
	&	$\dfrac{s_{\gamma}^{2}(s_{13}^{h})^{2}}{16\pi^{2}}
	\dfrac{m_{h}^{2}}{\tilde{M_{S}}}
	\ln\left( \dfrac{\tilde{M}_{S}^{2}}{m_{h}^{2}} \right)$
	&	$\dfrac{s_{\gamma}^{2}(s_{13}^{h})^{2}}{16\pi^{2}} {\tilde{M_{S}}}
	\ln\left( \dfrac{m_{h}^{2}}{\tilde{M}_{S}^{2}} \right)$	\\ \hline \hline
$N_{R}^{1}$
	&	$v_{\chi}-M_{N}^{ee}$
	&	$\dfrac{v_{\chi}^{2}}{M_{N}^{ee}}$	\\
$N_{R}^{2}$
	&	$\dfrac{3}{16\pi^{2}}\dfrac{M_{Z'}^{2}}{M_{N}^{ee}}
	\ln\left(\dfrac{M_{N}^{ee}}{M_{Z'}^{2}} \right)$
	&	$\dfrac{3}{16\pi^{2}}\dfrac{M_{Z'}^{4}}{(M_{N}^{ee})^{3}}
	\ln\left(\dfrac{M_{Z'}^{2}}{M_{N}^{ee}} \right)$	\\
$N_{R}^{3}$
	&	$v_{\chi}-M_{N}^{\tau\tau}$
	&	$\dfrac{v_{\chi}^{2}}{M_{N}^{\tau\tau}}$	\\ \hline \hline
$\mathcal{N}_{R}^{1}$
	&	$v_{\chi}+M_{N}^{ee}$
	&	$M_{N}^{ee}+\dfrac{v_{\chi}^{2}}{M_{N}^{ee}}$	\\
$\mathcal{N}_{R}^{3}$
	&	$v_{\chi}+M_{N}^{\tau\tau}$
	&	$M_{N}^{\tau\tau}+\dfrac{v_{\chi}^{2}}{M_{N}^{\tau\tau}}$ \\ \hline \hline 
\end{tabular}
\caption{Special cases given by the hierarchy between $v_{\chi}$ and $M_{N}^{ee}$. The left column resembles ISS while the right one the double SSM.}
\label{tab:Special-cases}
\end{table}

\section*{Conflicts of interests:}

The authors declare that we have no conflict of interest.

\printbibliography

\end{document}